\documentclass[twocolumn,showpacs,floatfix]{revtex4}
\usepackage{graphicx}
\usepackage{bm}

\begin{document}

\title{Role of transverse excitations in the instability of \\
       Bose-Einstein condensates moving in optical lattices}

\author{M. Modugno} \email{modugno@fi.infn.it}
\affiliation{%
\centerline{LENS - Dipartimento di Fisica, Universit\`a di Firenze and INFM 
  Via Nello Carrara 1, 50019 Sesto Fiorentino, Italy } and
  BEC-INFM  Trento, I-38050 Povo, Italy}

\author{C. Tozzo}
\affiliation{\centerline{Dipartimento di Fisica, Universit\`a di Trento, 
and Istituto Nazionale per la Fisica della Materia, BEC-INFM Trento,} 
I-38050 Povo, Italy }

\author{and F. Dalfovo}
\affiliation{\centerline{Dipartimento di Matematica e Fisica, Universit\`a
Cattolica, Via Musei 41, 25121 Brescia, Italy} and 
Istituto Nazionale per la Fisica della Materia, Unit\`a di Brescia
and BEC-INFM Trento, I-38050 Povo, Italy}

\date{\today}

\begin{abstract}
The occurrence of energetic and dynamical instabilities in a 
Bose-Einstein condensate moving in a one-dimensional (1D) optical 
lattice is analyzed by means of the Gross-Pitaevskii theory. Results 
of full 3D calculations are compared with those of an effective 1D 
model, the nonpolynomial Schr\"odinger equation, pointing out the 
role played by transverse degrees of freedom. The instability thresholds 
are shown to be scarcely affected by transverse excitations, so that
 they can be accurately predicted  by effective 1D models.
Conversely, transverse excitations turn out to be important
in characterizing the stability diagram and the occurrence of a complex 
radial dynamics above the threshold for dynamical instability. This 
analysis provides a realistic framework to discuss the dissipative 
dynamics observed in recent experiments.

\end{abstract}
\maketitle

\section{Introduction}

The occurrence of energetic and dynamical instabilities in a
Bose-Einstein condensate (BEC) that moves in a periodic (optical)
potential is an interesting problem from the conceptual viewpoint,
since it involves basic properties of superfluids. Experimentally the
dissipative dynamics of BECs moving through optical lattices has been
investigated both in the weak \cite{burger,burger2} and tight-binding
regimes \cite{cataliotti,lens}, and a number of papers have recently
been published in connection with these and similar experiments
\cite{morsch,oberthaler,smerzi,menotti,pethick,wuniu,adhikari,nesi,taylor}.

 From the theoretical side, systematic investigations of the stability
regimes have been so far presented for strictly one-dimensional (1D)
systems. For deep optical lattices (tight-binding limit) it has been
argued that the major mechanism responsible for the superflow
breakdown of a moving BEC is the onset of a dynamical
(\textit{modulational}) instability \cite{wuniu,smerzi}. This behavior
has also been confirmed by direct integration of the 3D
Gross-Pitaevskii (GP) equation in the same regime
\cite{adhikari,nesi}, and well reproduces the experimental
observations \cite{lens}.  Conversely, the opposite limit of shallow
lattices has been the object of a stimulating debate and it has been
suggested that the origin of the observed instability could have an
energetic or dynamic character \cite{burger,burger2,wuniu}.

Here we present a general discussion of the energetic and dynamical
instabilities within the GP theory. We make use of a 3D description
that includes the radial degrees of freedom in order to provide a
reliable framework for a comparison with current experiments. In
Sec. II, we briefly summarize the formalism of the standard linear
analysis within the GP theory. Then, in Secs. III and IV, we discuss
in detail the case of a cylindrical condensate in a 1D optical
lattice, for which one can use Bloch functions and rigorously define
the concept of quasimomentum associated with the motion of the
condensate in the lattice. For any given quasimomentum of the
condensate, we calculate the excitation spectrum and the stability
diagram by solving the equations of the linearized GP theory
\cite{review}. The same quantities are also calculated by means of an
effective 1D model, known as nonpolynomial Schr\"odinger equation
(NPSE) \cite{salasnich}, which includes the transverse direction
through a Gaussian ansatz for the radial shape of the order parameter,
with a $z$- and $t$-dependent width. With respect to strictly 1D
models, which rely on a suitable renormalization of the mean-field
coupling constant, the NPSE has the advantage that the true 3D
coupling constant $g=4\pi \hbar^2 a/m$ can be used, thus allowing for
a direct comparison with experiments. It turns out that this model
gives accurate predictions for the instability thresholds, the latter
being mainly determined by the dispersion of the lowest branch of
excitations, with no radial nodes. Higher branches, with one or more
radial nodes, that are included in the GP theory but not in the
NPSE, are shown to be important in characterizing the number and the
type of excitations that become unstable above the instability
threshold. They also produce a significant change in the shape of the
dynamically unstable regions above threshold.

Finally, in Sec. V, we perform a GP simulation for the elongated
condensate of the experiments of Ref. \cite{burger}. The results of
the simulation are analyzed by using the spectra and the phase
diagrams of the cylindrical condensate of Secs. III and IV. This
analysis provides convincing evidence that the breakdown of superfluid
flow observed in Ref. \cite{burger} is associated with the onset of a
dynamical instability, triggered by the resonant coupling of
Bogoliubov phonon and antiphonon modes. The subsequent dynamics is
shown to strongly involve radial motions.

\section{GP equation and linear analysis}
\label{sec:gpe}

The GP equation for the order parameter $\psi$ of $N$ condensed 
atoms of mass $m$ and scattering length $a$ is \cite{review}
\begin{equation}
i\hbar \frac{\partial}{\partial t}\psi(\mathbf{x},t)=
\biggl[-\frac{\hbar^2}{2m}\nabla^2+V(\mathbf{x})+
g|\psi(\mathbf{x},t)|^2\biggr]\psi(\mathbf{x},t)\,,
\label{eq:GP}
\end{equation}
where $g=4\pi\hbar^2a/m$. We will consider the case of condensates 
confined in a harmonic potential superimposed to a 1D periodic 
optical lattice, 
$V(\mathbf{x})= V_{\rm ho}(\mathbf{x}) + V_{\rm L}(z)$. 

Let $\psi_0({\bf x},t)=\phi_{0}({\bf x})\exp(-i\mu t/\hbar)$ be a 
solution of the stationary GP equation, having chemical potential 
$\mu$,
\begin{equation}
\biggl[-\frac{\hbar^2}{2m}\nabla^2+V(\mathbf{x})+
g|\phi_{0}({\bf x})|^2\biggr]\phi_{0}({\bf x}) = \mu \phi_{0}({\bf x})\,,
\label{eq:stationaryGP}
\end{equation}
and  normalized according to $\int \! d^{3}x \ |\phi_{0}({\bf x})|^2
= N$.  If the state $\phi_0$ is a local minimum of the energy functional
\begin{equation}
E[\phi_0]=\int \! d^{3}x \ 
\phi_{0}^{*}\left[-\frac{\hbar^2}{2m}\nabla^2 +V -\mu\right]\phi_{0}
 +\frac{g}{2}|\phi_{0}|^4 
\label{eq:gpener}
\end{equation}
the system is energetically stable.  Energetic (Landau) instability
sets in when this condition is no longer true, and there are some
directions in the $\psi$ space along which it is possible to lower the
energy. This happens, for instance, when a uniform condensate flows
with velocity $v$ in presence of a static external potential; if $v$
is larger than the velocity of Bogoliubov's sound, then the system can
lower its energy by emitting phonons (Landau criterion of
superfluidity \cite{landau}). Similar conditions can be found for an
inhomogeneous condensate moving in an optical lattice above a certain
critical velocity $v_E$.

The condition for energetic stability can be formally derived by 
considering small deviations from $\phi_0$ in the form 
\begin{equation}
    \psi({\bf x},t)= 
    e^{\displaystyle i\mu t/\hbar} [\phi_0({\bf x}) 
+ \delta\phi({\bf x},t) ] \, , 
    \label{eq:fluctuations}
\end{equation}
and expanding the energy functional up to the quadratic terms
\begin{equation}
 E[\phi]=E[\phi_{0}]+\int \! d^{3}x \ 
\left({\delta\phi^*, \delta\phi}\right)M
    \pmatrix{{\delta\phi}\cr{\delta\phi^*}\cr} \, , 
\end{equation}
where 
\begin{equation}
    M=\pmatrix{H_0+2g|\phi_0|^2 &g\phi_0^2\cr
    g\phi_0^{*2}&H_0+2g|\phi_0|^2\cr}
\end{equation}
and
\begin{equation}
H_{0}=\left[-\frac{\hbar^2}{2m}\nabla^2+V-\mu\right].
    \label{eq:h0}
\end{equation}
The onset of energetic (Landau) instability is thus signaled by the 
appearance of negative eigenvalues in the spectrum of the operator 
$M$. 

A second type of instability is the dynamical (modulational) 
instability. This occurs when the frequency of some modes in the
excitation spectrum has a nonzero imaginary part. Then the occupation 
of these modes grows exponentially in time and rapidly drives the 
system away from the steady state.  The conditions for the onset 
of this instability can be obtained again by means of the expansion 
(\ref{eq:fluctuations}). When inserted into the GP equation 
(\ref{eq:GP}) it yields the Bogoliubov equations
\begin{equation}
    \sigma_z M \pmatrix{{\delta\phi}\cr{\delta\phi^*}\cr} =
    \hbar \omega \pmatrix{\delta\phi\cr\delta\phi^*\cr}
\end{equation}
whose character is now determined by the operator $\sigma_z M$, with
\begin{equation}
    \sigma_z\equiv \pmatrix{1&0\cr0&-1\cr}.
\end{equation}

Energetic and dynamical instabilities are therefore related to 
the spectral properties of the operator $M$ and $\sigma_z M$ 
\cite{wuniu,castin}. This has two important consequences: 
(i) when the system is energetically stable it is also dynamically 
stable; 
(ii) the energetic instability, which is related to the 
spectrum of $M$, cannot be revealed by a direct integration of 
the time-dependent GP equation (\ref{eq:GP}), which is 
dissipationless and is governed by the spectrum of $\sigma_z M$.

\section{Spectrum of a cylindrical condensate in a lattice}
\label{sec:spectrum}

In order to perform explicit calculations, let us consider the case 
of an infinite cylindrical condensate, which is radially confined by
the harmonic potential $V_{\rm ho} (r) = (1/2) m \omega_r^2 r^2$, 
and is subject to the periodic potential $V_{\rm L} (z) = sE_R 
\cos^2 (q_B z)$. The Bragg wave vector $q_B=\pi/d$ is determined 
by the lattice spacing $d$. The quantity $E_{\rm R}\equiv \hbar^2 
q_B^2/2m$ is the recoil energy of an atom absorbing one lattice photon 
and $s$ is a dimensionless parameter fixing the lattice intensity. 
Such a cylindrical condensate is an accurate representation of an 
elongated axially symmetric condensate, 
when the weak axial harmonic confinement can be neglected. The major 
advantage of this geometry is that one can exploit the periodicity 
of the system by expanding the order parameter on a basis of Bloch 
functions, with the same period of the lattice. 

\subsection{Linearized GP equation}

Within the GP theory, the starting point of the linear stability 
analysis for the cylindrical condensate is the expression 
\begin{equation}
\psi(r,z,t)= {\rm e}^{\displaystyle -i\mu_p t} 
{\rm e}^{\displaystyle
ipz} \left[\phi_{p0}(r,z) +\delta\phi_{p}(r,z,t)\right] \,. 
\end{equation}
The Bloch wave vector $p$ (also called quasimomentum of the 
condensate) is associated with the velocity of the condensate 
in the lattice and is restricted to the first Brillouin zone 
\cite{wuniu,menotti}. For a given $p$, the function $\phi_{p0}$
is the solution of the stationary GP equation with energy $E_p$,
while the excited part of the order parameter can be 
written as  
\begin{eqnarray}
\delta\phi_{p} (r,z,t)&=&\sum_{q,j}\left( u_{pq,j}(r,z)
{\rm e}^{\displaystyle i(qz-\omega_{pq,j}t)} \right.  \nonumber\\
&& \left.+ v^*_{pq,j}(r,z){\rm e}^{\displaystyle
-i(qz-\omega_{pq,j}t)}\right) \,,
\end{eqnarray}
where $q$ is the Bloch wave vector (or quasimomentum) of the 
excitations, and $j$ represents all possible values of the two 
quantum numbers $\alpha=1,2,3,\dots$ and $\nu=0,1,2,\dots$.  The 
former is Bloch band index and the latter is the number of radial 
nodes in the Bogoliubov quasiparticle amplitudes $u$ and $v$. Along 
$z$ all the functions $\phi_{p0}$, $u_{pq,j}$ and $v_{pq,j}$ are 
periodic with the same period of the lattice. 

Inserting the above expressions into the GP equation (\ref{eq:GP}), 
one gets the Bogoliubov equations 
\begin{eqnarray}
&&\hspace{-1cm}\left[H_{r}+H_{z+}-\mu_p + 2g|\phi_{p0}|^2\right]u+
g\phi_{p0}^2 v =\hbar\omega u
\label{eq:bogol1}
\\
&&\hspace{-1cm}\left[H_{r}+H_{z-}-\mu_p + 2g|\phi_{p0}|^2\right]v+
g\phi_{p0}^{*2} u =-\hbar\omega v
\label{eq:bogol2}
\end{eqnarray}
with
\begin{eqnarray}
H_{r}&=&-{\hbar^2\over 2m}\nabla_r^2+\frac12 m
\omega_r^2r^2\\
H_{z\pm}&=&{\hbar^2(-i\partial_z\pm p+q)^2\over
2m}+ s E_R \cos^2(q_Bz) \,. 
\end{eqnarray}
The quasiparticle amplitudes $u$ and $v$ satisfy the relation 
\begin{equation}
\int \! d^2r \int^{d/2}_{-d/2} \! dz \ (u_{pq,i}u_{pq,j}^* - 
v_{pq,i} v_{pq,j}^*) = \delta_{ij} \,. 
\label{eq:ortho}
\end{equation}
Note that, for each quasiparticle obeying this orthonormality 
conditions, the Bogoliubov equations admit an antiquasiparticle 
with norm $-1$, which corresponds to replacing
the amplitudes $(u,v)$ with $(v^*,u^*)$  \cite{wuniu,castin}. 

The above equations can be solved numerically. We choose the 
relevant parameters such to simulate typical elongated condensates 
of the experiments of Ref. \cite{burger}. The linear atomic density 
in the $z$ direction, averaged over the lattice length $d$, is 
taken to be equal to the linear density of the actual 3D condensate, 
close to the trap center and in the absence of the  lattice.
Here the chemical potential for $p=0$ is taken to be 
$\mu_0=7.2~\hbar \omega_r$. We also use $2\pi \ q_B^{-1} = 800$~nm 
and $\omega_r=2\pi \times 90$~Hz. For each $p$, the order parameter 
$\phi_{p0}(r,z)$ and its energy $E_p$ are calculated by solving the
stationary GP equation by means of iterative methods. Then the
Bogoliubov equations (\ref{eq:bogol1}) and (\ref{eq:bogol2}) are 
transformed into matrix equations by projecting the functions 
$u$ and $v$ on the Bessel-Fourier basis $|l,n\rangle\equiv
J_0\left({\alpha_{0n}r}/{r_{max}}\right) \exp(i2q_Blz)$, where 
$\alpha_{0n}$ are zeros of the Bessel functions $J_0(r)$ and
$r_{max}$ is the radius of the computational box \cite{tozzo,iyanaga}. 
The resulting equations are not diagonal over the Fourier index $l$,
and involve the numerical diagonalization of a $2\times N_r 
\times N_z$ rank matrix.

\begin{figure}
\centerline{\includegraphics[width=7.7cm,clip=]{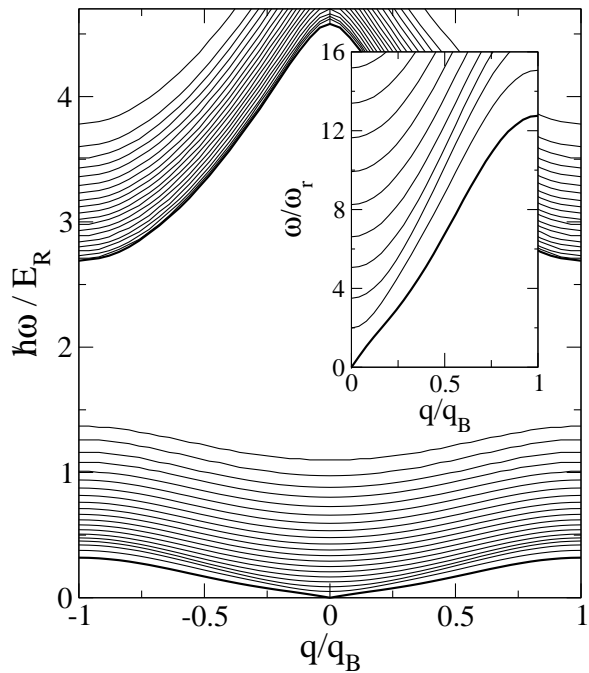}}
\caption{Excitation spectrum of a cylindrical condensate
at rest in the optical lattice (quasimomentum of the condensate 
$p=0$) as a function of the quasimomentum $q$ of the excitations, 
for lattice intensity $s=5$. The first two Bloch bands are shown and 
for each band we plot the first $20$ radial branches. The lowest 
branch in each band (thickest line) corresponds to axial excitations 
with no radial nodes. In the inset we show a magnification of the low
energy spectrum. }
\label{fig:spectrum-p0}
\end{figure}

\begin{figure}
\centerline{\includegraphics[width=7.7cm,clip=]{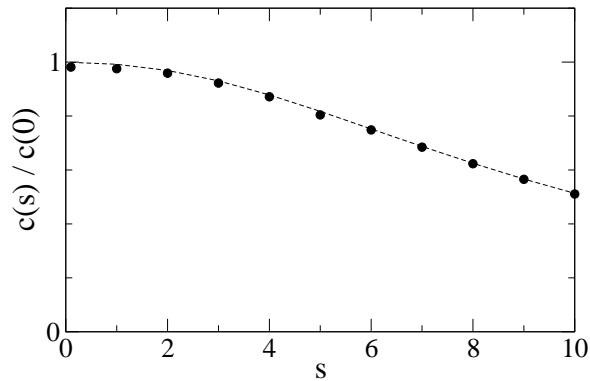}}
\caption{Bogoliubov sound velocity $c$ as a function of the 
lattice intensity $s$. Points: slope of the lowest phononic 
branch in the Bogoliubov spectrum of the cylindrical 
condensate. Dashed line: analytic prediction 
$c=(\kappa m^*)^{-1/2}$ (see text).   }
\label{fig:sound}
\end{figure}

The $p=0$ case corresponds to the condensate at rest. In 
Fig.~\ref{fig:spectrum-p0} we show a typical spectrum, for $s=5$. 
The first two Bloch bands ($\alpha=1$ and $2$) are shown. For 
each band, we plot the first $20$ radial branches ($\nu=0,1,2,
\dots,19$). The thickest lines are the dispersion laws of 
the $\nu=0$ excitations. A magnification of the low-energy 
branches is shown in the inset. The lowest branch corresponds 
to the dispersion law of $\nu=0$ axial phonons. It starts 
linearly at $q \to 0$ and its slope is the Bogoliubov sound 
velocity $c$. We calculate the slope for different values of 
the lattice intensity $s$ and we plot the results in 
Fig.~\ref{fig:sound} (points). The dashed line is the analytic 
prediction $c=(m^* \kappa)^{-1/2}$ of Ref.~\cite{meret}, where 
the effective mass $m^*$ and the compressibility $\kappa$ are 
defined as $m^* = \lim_{p\to 0} (\partial^2E_p/\partial p^2)^{-1}$ 
and $\kappa^{-1}= \bar{n} (\partial \mu/\partial \bar{n})$ 
(see also the discussion in Ref. \cite{taylor}). Here we calculate 
the effective mass from the numerical values of the energy $E_p$.
The compressibility is taken to be the one of a uniform gas 
in the same 1D optical lattice and with density equal to the 
average density of the cylindrical condensate (for the radial average
we use the Thomas-Fermi value $\bar{n}=n(0)/2$ where $n(0)$ is the
central density). The agreement is remarkable.  The value at $s=0$ is
the sound velocity $c= [g n(0)/(2m)]^{1/2}$ in a cylindrical
condensate without lattice \cite{zaremba}. Finally, we note that the
$\nu=1$ branch in the inset of Fig.~\ref{fig:spectrum-p0} exactly
starts at $\omega=2\omega_r$ for $q=0$, where it corresponds to the
radial breathing mode of the cylindrical condensate.

\begin{figure}
\centerline{\includegraphics[width=7.7cm,clip=]{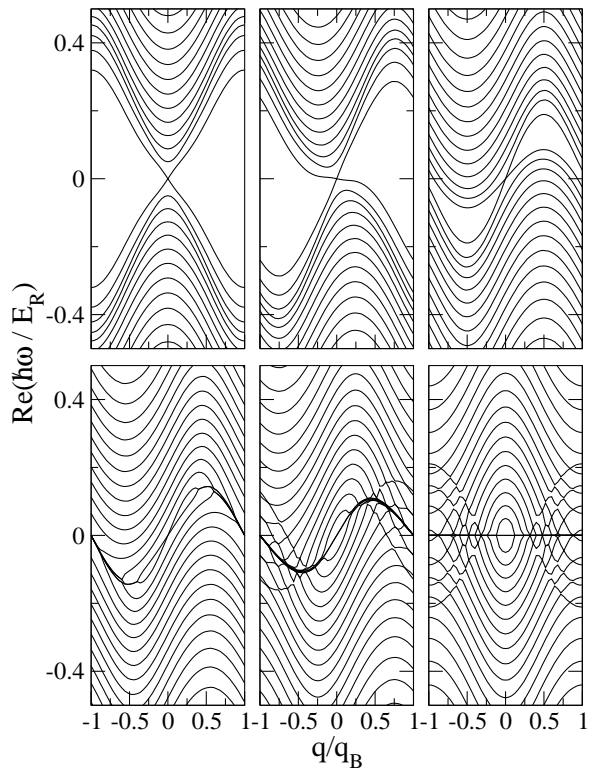}}
\caption{Real part of the excitation spectrum of a cylindrical 
condensate in a lattice with $s=5$ and for different values of 
the condensate quasimomentum $p$ ($p/q_B=0,0.25,0.5,0.55,0.75,1$), 
calculated from Eqs.~(\protect\ref{eq:bogol1})-(\protect\ref{eq:bogol2}).}
\label{fig:spectraGP}
\end{figure}

At $p\neq 0$, when the condensate moves in the lattice, the whole 
spectrum changes. In Fig.~\ref{fig:spectraGP} we show a sequence
of six spectra for $s=5$ and $p/q_B=0, 0.25, 0.5, 0.55, 0.75, 1$. 
The real part of the frequencies $\omega_{pq,j}$ is plotted. At 
$p=0$ quasiparticles and antiquasiparticles have the same energy, 
since the system is symmetric under $z \to -z$ inversion. For a 
given $q$, they correspond to excitations propagating in opposite 
directions (opposite sign of $\omega$). In Fig.~\ref{fig:spectraGP} 
the spectrum of antiquasiparticles, at $p=0$, is plotted in the 
negative $\omega$ plane. 
 
By increasing $p$, the slope of the $\nu=0$ phononic branch 
increases for phonons which propagate forwards on top 
of the traveling condensate. Vice versa it decreases for those 
propagating backwards and their dispersion law eventually 
crosses the $\omega=0$ axis and changes sign. By further increasing 
$p$, one notices that the frequency of both phonons and antiphonons 
approaches zero at the boundaries of the Brillouin zone.
At a critical quasimomentum $p_D$ ($p_D=0.525~q_B$ in our case), the 
real part of the $\nu=0$ mode frequency vanishes at $q=\pm q_B$. 
Phonon and antiphonon pairs thus exhibit a resonance coupling
by first-order Bragg scattering, giving rise to a dynamical 
instability. For 1D condensates this process has been already 
discussed in Ref. \cite{wuniu}. The coupled pairs of excitations, 
with $|q|=q_B=\pi/d$, start having imaginary frequency. The system 
can therefore develop a macroscopic standing wave, whose wavelength 
is twice the lattice spacing, which implies an out-of-phase 
oscillation of the condensate population in adjacent lattice sites. 
This is also consistent with the results of the 1D simulations of 
Ref. \cite{menotti}. 

\begin{figure}
\centerline{\includegraphics[width=7.7cm,clip=]{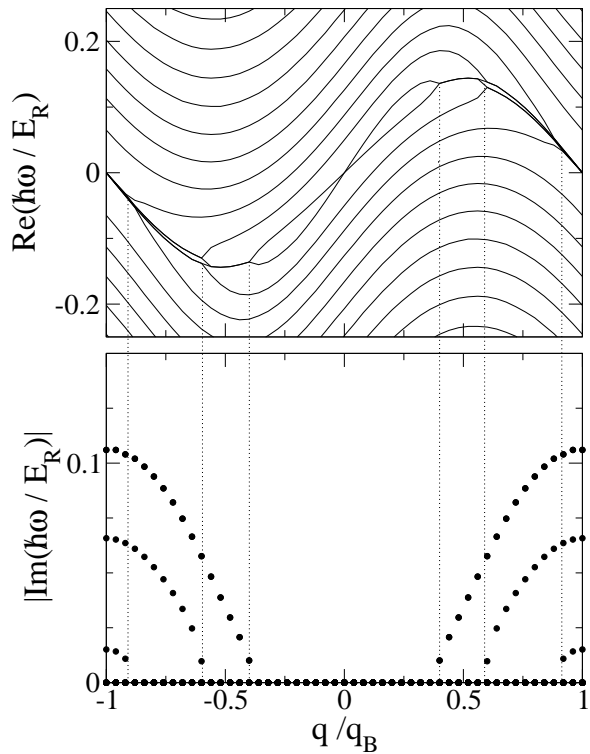}}
\caption{Real (top) and imaginary (bottom) parts of the frequencies 
obtained from Eqs.~(\protect\ref{eq:bogol1})-(\protect\ref{eq:bogol2})
for $s=5$ and $p/q_B=0.55$. The vertical (dotted) lines correspond to 
the points where a conjugate pair of complex frequencies appears. }
\label{fig:spectrumGP-055}
\end{figure}

For $p>p_D$ the coupling between $\nu=0$ phonons and antiphonons 
extends from zone boundary to lower values of $q$. Moreover, modes 
with radial nodes ($\nu>0$) also couple, first at $q=q_B$ and then 
down to lower $q$'s. A conjugate pair of complex frequencies appears 
each time a resonant coupling occurs between a pair of quasiparticle 
and antiquasiparticle that are degenerate prior to the coupling. 
An example is given in Fig.~\ref{fig:spectrumGP-055} where we plot 
the real and imaginary parts of the eigenfrequencies obtained by 
solving Eqs.(\ref{eq:bogol1}) and (\ref{eq:bogol2}) in the case 
$p/q_B=0.55$. Finally, as one can see in the last two frames of 
Fig.~\ref{fig:spectraGP}, a complicate sequence of no-crossing patterns 
also develop between modes belonging to the quasiparticle spectrum 
or between those belonging to the antiquasiparticle spectrum. These 
no-crossings do not produce any dynamical instability, but they 
contribute to mix up several radial branches {\it via} hybridization 
processes, which are not included in 1D models.  

\subsection{Nonpolynomial Schr\"odinger equation}

The description of the system can be simplified by using an 
effective 1D model, the nonpolynomial Schr\"odinger equation (NPSE) 
\cite{salasnich}, which partially includes also the radial to axial 
coupling, and has been shown to provide a realistic description in 
several situations \cite{salasnich,massignan,nesi}.  This model is 
obtained from the GP equation by means of a factorization of the 
order parameter in the product of a Gaussian radial component of $z$-
and $t$-dependent width, $\sigma(z,t)$, and of an axial wave
function  $\psi(z,t)$ that satisfies the differential equation
\begin{eqnarray}
\label{eq:npse-1}
i\hbar\frac{\partial}{\partial t}\psi(z,t)&=&
\bigg[-\frac{\hbar^2}{2m}\nabla_z^2+V(z)+\frac{gN}{2\pi\sigma^2}
|\psi|^2\\ &&+\frac{1}{2}\hbar
\omega_r\left(\frac{a_r^2}{\sigma^2}+
\frac{\sigma^2}{a_r^2}\right)\bigg]\psi(z,t)\,, \nonumber
\end{eqnarray}
coupled with an algebraic equation for the radial width
\begin{equation}
\sigma(z,t)=a_r\sqrt[4]{1+2aN|\psi(z,t)|^2} \, . 
\label{eq:npse-2}
\end{equation}
Here $a_r=\sqrt{\hbar/m\omega_r}$ is the radial oscillator length.
By expanding $\psi(z,t)$ around the stationary wave function 
$\phi_0(z)$ and using Eqs.~(\ref{eq:npse-1}) and (\ref{eq:npse-2}), 
it is straightforward to obtain the Bogoliubov-like equations
\begin{equation}
\pmatrix{H_0+A &B\cr-B^*&-(H_0+A)\cr}
\pmatrix{\delta\phi\cr\delta\phi^*\cr}=\hbar\omega
\pmatrix{\delta\phi\cr\delta\phi^*\cr}\,,
\label{eq:bog-NPSE}
\end{equation}
\begin{eqnarray}
H_0 &=& -\frac{\hbar^2}{2m}\nabla_z^2 + V(z)\\ 
A &=&
\frac{gN}{2\pi}\frac{|{\phi_0}|^2}{\sigma^2}\left( 2-Na|\phi_0|^2
\frac{a_r^4}{\sigma^4}\right)\\ 
&+& \frac12\hbar\omega_r\left[ \frac{\sigma^2}{a_r^2}
+\frac{a_r^2}{\sigma^2}+Na\phi_0^2\frac{a_r^2}{\sigma^2}\left(
1-\frac{a_r^4}{\sigma^4}\right)\right]\,, \nonumber\\ 
B &=&
\frac{gN}{4\pi}\frac{{\phi_0}^2}{\sigma^2}\!\left[
\frac52 - 2Na|{\phi_0}|^2\frac{a_r^2}{ \sigma^2}
-\frac12\frac{a_r^4}{\sigma^4}
\right]\,.
\end{eqnarray}
The solution of these equations is obtained by a numerical 
matrix diagonalization. Compared to the solution of the Bogoliubov 
equations (\ref{eq:bogol1}) and (\ref{eq:bogol2}), the calculation
of the spectrum from Eqs.~(\ref{eq:bog-NPSE}) is much faster. 

Owing to the partial coupling between axial and radial degrees of
freedom, the NPSE provides a more accurate description of the 
actual 3D condensates with respect to strictly 1D models. An
important advantage consists in the fact that, differently from 
strictly 1D models, the NPSE does not require any renormalization 
procedure for the coupling constant $g$, thus offering the 
possibility to make a direct comparison with experiments, as
well as with GP calculations for 3D condensates. In this respect 
one has to remark also that, since the radial degrees of freedom 
are included through a single function $\sigma(z,t)$, the NPSE 
only accounts for $\nu=0$ radial excitations. The comparison 
between the predictions of the NPSE and those of the GP equation
for the cylindrical condensate is thus particularly helpful
to point out the role of the $\nu>0$ radial modes. 

\begin{figure}
\centerline{\includegraphics[width=7.7cm,clip=]{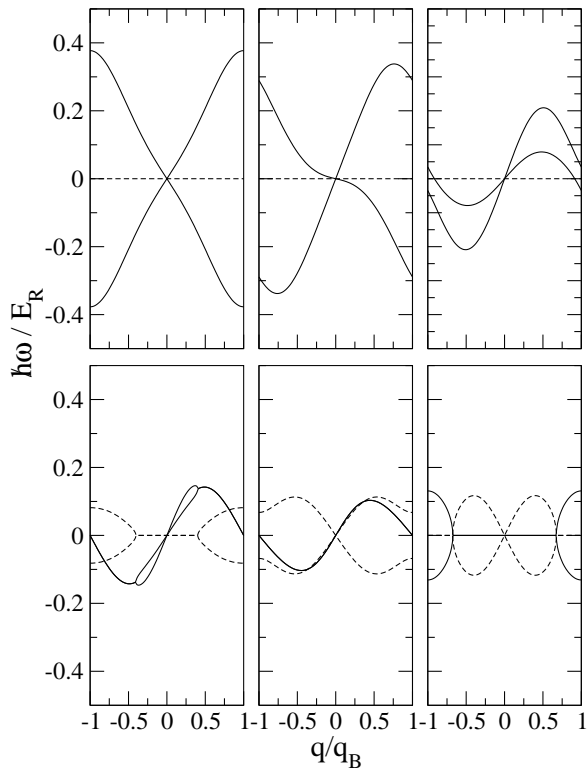}}
\caption{Excitation spectrum obtained from the NPSE for a cylindrical 
condensate in a lattice with $s=5$ and for different values of the 
quasimomentum $p$, as in Fig.~\ref{fig:spectraGP} 
($p/q_B=0,0.25,0.5,0.55,0.75,1$). }
\label{fig:spectraNPSE}
\end{figure}

In Fig.~\ref{fig:spectraNPSE} we plot the real (solid lines) and
imaginary part (dashed lines) of the excitation
frequencies obtained with the NPSE for the same $s$ and $p$ of
Fig.~\ref{fig:spectraGP}. The two branches correspond to the $\nu=0$
phonon and antiphonon dispersions and their behavior is shown to
closely follow that of the lowest $\nu=0$ branches of the GP case. In
particular, the critical value $p=p_D$ at which the two modes couple
at $q=q_B$, giving rise to dynamical instability, coincides with the
GP result ($p_D=0.525~q_B$). The qualitative behavior is very similar
also to that found in the 1D analysis of Ref. \cite{menotti}.

\section{Stability diagrams}
\label{sec:stability}

\begin{figure}
\centerline{\includegraphics[width=7.7cm,clip=]{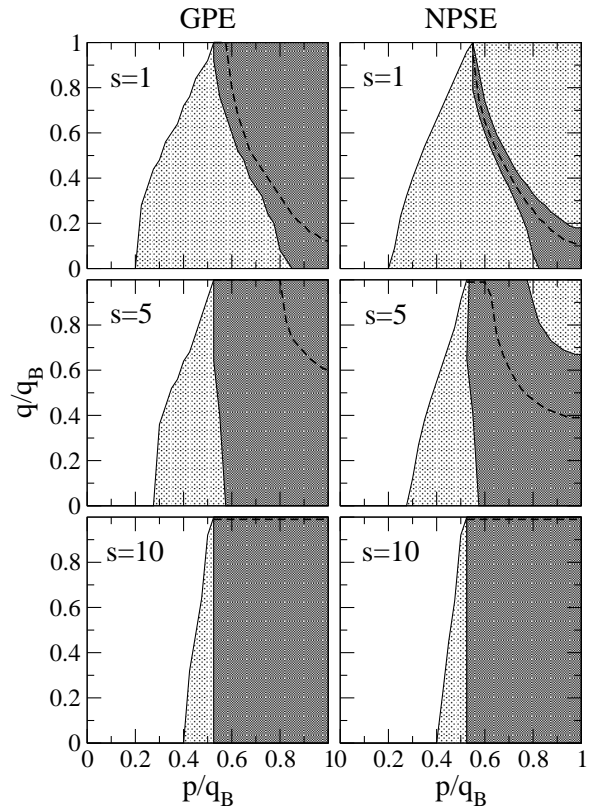}}
\caption{Stability diagrams obtained from the GP spectra (left)
and NPSE spectra (right) as a function of the quasimomenta $p$ 
and $q$ of the condensate and of the quasiparticles, respectively, 
and for $s= 1, 5, 10$. Shaded areas represent the regions where 
the system is dynamically (dark) and/or energetically (both 
light and dark) unstable. Dashed lines correspond to the modes
having the largest imaginary part (most unstable modes). }
\label{fig:stability}
\end{figure}

By repeating the calculations of the Bogoliubov frequencies
(eigenvalues of the matrix $\sigma_z M$) for different values 
of the condensate quasimomentum $p$, one can draw the dynamically 
unstable regions in the $p$-$q$ plane. In the same way, one can 
find the spectrum of the matrix $M$ and draw the regions of
energetic instability. The calculations can be performed both
with GP theory and NPSE. Typical results are shown in 
Fig.~\ref{fig:stability} for $s=1, 5,$ and $10$. The 
white area corresponds to a stable condensate;
the light shaded area to a condensate which is energetically 
unstable (in the presence of dissipative processes, the energy 
can be lowered by emitting phonons of quasimomentum $q$ which 
lie in the shaded range); finally the dark shaded area is the
region where the system is both energetically and dynamically 
unstable. The thresholds for energetic and dynamical instability 
correspond to the lowest values of $p$ at which the instability 
occurs. Let us call them $p_E$ and $p_D$, respectively.     

Two main features emerge from the comparison of GP with NPSE: 
(i) the left border of the energetically and dynamically unstable 
regions are very similar in the two cases, which also implies a
a good agreement for the critical quasimomenta $p_E$ and $p_D$; 
(ii) for small values of $s$ the shape of the dynamically unstable 
region above threshold looks rather different. 

\begin{figure}
\centerline{\includegraphics[width=7.7cm,clip=]{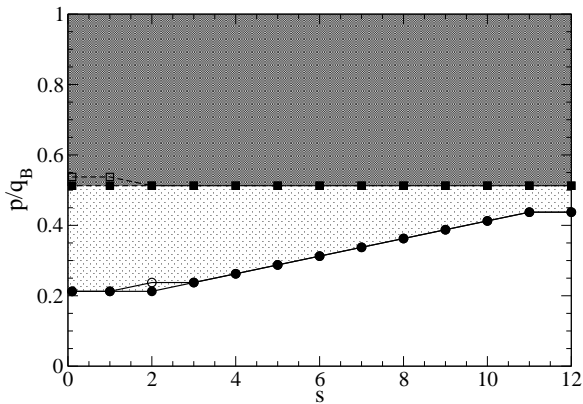}}
\caption{Thresholds for energetic (circles) and dynamical (squares)
instability as a function of the lattice intensity $s$.  The results
of GP (filled symbols) and NPSE (empty symbols) are almost indistinguishable. 
Shaded areas represent the regions where the system is dynamically (dark)
and/or energetically (both light and dark) unstable, according to the
GP. }
\label{fig:thresholds}
\end{figure}

The first point is better visualized in Fig.~\ref{fig:thresholds},
where we show the results for $p_E$ and $p_D$ as a function of $s$.
The predictions of the two models are almost indistinguishable for
both the energetic and dynamical instabilities in the whole range of
$s$ here considered. These results confirm that the NPSE can be used
as an efficient model to quantitatively identify the onset of
instability for a wide range of $s$, relevant for available
experiments \cite{nesi}.  The reason for its accuracy in predicting
the thresholds is simply that the critical quasimomenta $p_E$ and
$p_D$ are determined by the behavior of the $\nu=0$ branch of
excitations with no radial nodes. A comparison between
Figs.~\ref{fig:spectraGP} and \ref{fig:spectraNPSE} shows that these
modes are indeed very accurately described by the NPSE up to the
instability threshold.

A closer inspection of the stability diagrams shows a difference 
in the shape and size of the dynamically unstable region at small
$s$. The one predicted by NPSE does not fill the whole region above 
threshold. This agrees with previous 1D models \cite{wuniu,menotti}, 
but disagrees with the predictions of 3D GP equation. This 
disagreement can be easily understood by looking at the role
played by the transverse degrees of freedom. In the NPSE, which only
accounts for $\nu=0$ modes, nonvanishing complex frequencies appear
when phonons and antiphonons couple, collapsing onto a single 
dispersion law. For a given $p$, this provides a range of $q$ for
unstable modes, which starts first at $q_B$ and then  lowers 
towards $q\to 0$ (see Fig.~\ref{fig:spectraNPSE}). However, at high
enough $p$, the dispersion of the phonon-antiphonon pair splits
again into two separate curves and this gives a window of 
stable modes from a certain $q$ (at fixed $p$) up to zone boundary. 
This window is the dynamically stable region on the top-right 
corner of the NPSE stability diagrams in Fig.~\ref{fig:stability}.
Now, by looking at the GP spectra in Fig.~\ref{fig:spectraGP}, one can
easily understand why these stable regions are absent in the GP 
stability diagrams. The $\nu=0$ modes in this case are mixed up 
with several $\nu > 0$ branches and these radial modes can be 
unstable also in the range of $q$ where the $\nu=0$ modes are 
stable. It is worth stressing that the occurrence of 
complex frequencies for such radial excitations also implies
that, once the system is brought into the unstable region 
above $p_D$, it can easily develop macroscopic motions with
a nontrivial radial dynamics.  

\begin{figure}
\centerline{\includegraphics[width=7.7cm,clip=]{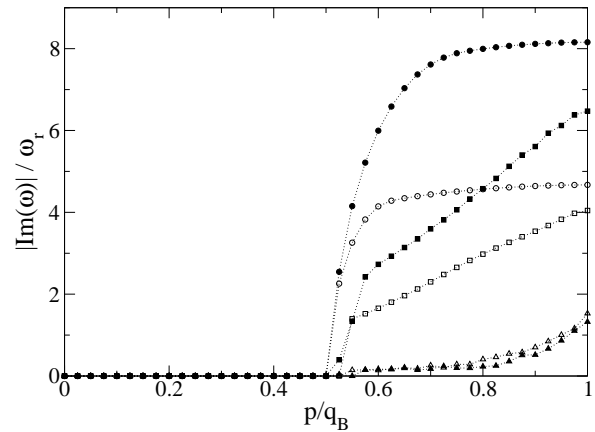}}
\caption{Growth rates of the most dynamically unstable modes for 
different values of the lattice intensity $s$ ($s=0.1$ triangles,
$s=1$ squares, $s=5$ circles) obtained with both GP and NPSE (filled 
and empty symbols, respectively).}
\label{fig:growth-rate}
\end{figure}

Finally, it is interesting to compare the predictions of GP and 
NPSE for the growth rate, $|{\rm Im} (\omega) |$, of the most dynamically 
unstable modes. The results are shown in Fig.~\ref{fig:growth-rate}. 
At small $s$, the agreement between the two calculations (solid and
empty symbols) is rather good. By increasing $s$ the agreement 
for the absolute values of growth rates becomes less and less 
quantitative. However, for each value of $s$, the shape of the GP 
and NPSE curves is still quite similar. This result is interesting
in view of the comparison with experiments, where the shapes of 
these curves can be more relevant than their absolute values.
In the experiments, in fact, it is much more likely to measure loss 
rates (or lifetimes) on large timescales, and estimate their relative
increase as a function of $p$, rather than the absolute values of
the growth rates in the small amplitude linear regime (short times). 

\section{An experiment revisited}

In this section we compare the predictions of the linear analysis 
discussed so far with the direct solutions of the time-dependent 
Gross-Pitaevskii equation (\ref{eq:GP}). As an example we consider 
the case of the LENS experiment presented in Ref. \cite{burger}, which has 
been the object of a stimulating debate on the origin of the observed 
instability \cite{burger2}. Indeed two interesting questions have 
been raised: (i) whether the GP equation is suitable to describe the 
phenomenology observed in the experiment, and (ii) whether the
observed breakdown of superfluid flow comes from energetic and/or 
dynamical instability.

To model the LENS experiment \cite{burger}, we consider an elongated 
$^{87}\mathrm{Rb}$ condensate with $N=3\times 10^5 $ atoms confined 
in an axially symmetric harmonic trap of frequencies $\omega_r = 2\pi
\times 90$ Hz, $\omega_z = 2\pi \times 8.7$ Hz. 
The optical lattice along $z$ is 
produced with laser beams of wavelength  $\lambda= 2 \pi/q_B = 795 $~nm 
and intensity $s=1.59$. The system is prepared in the ground state of 
the combined harmonic+periodic potential and then is let evolve 
after a sudden displacement $\Delta z$ of the harmonic trapping.

\begin{figure} 
\centerline{\includegraphics[width=7.7cm,clip=]{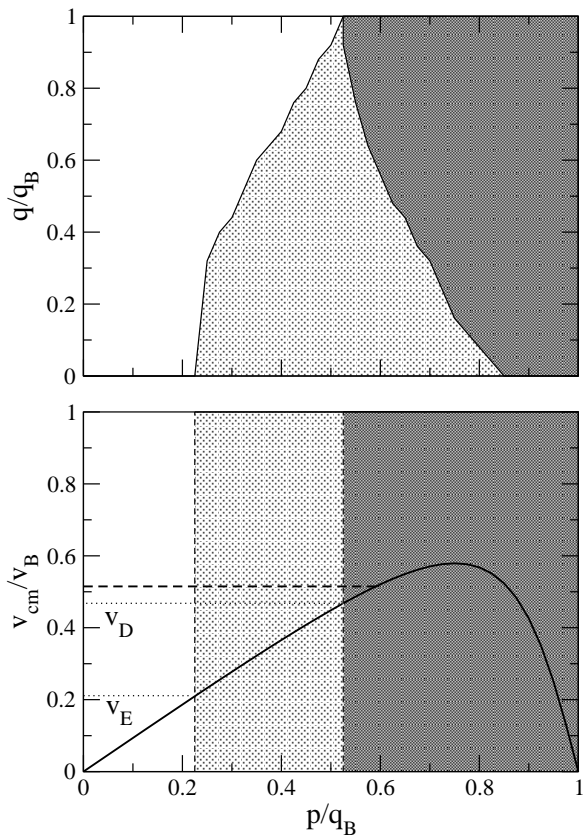}}
\caption{Top: GP stability diagram for a cylindrical condensate
which approximates the experimental elongated condensate of 
Ref.~\protect\cite{burger} in its central region. Light and dark 
shaded areas are unstable regions, as in 
Fig.~\protect\ref{fig:stability}. Bottom: 
center-of-mass velocity as a function of the condensate 
quasimomentum ($v_B=\hbar q_B/m$). 
The critical velocities for the energetic ($v_E$) and
dynamical ($v_D$) instabilities of the cylindrical condensate
are shown as horizontal dotted lines. The horizontal dashed line
is the velocity at which the instability occurs in the GP simulation
of the experimental 3D condensate of Ref.~\protect\cite{burger}. }
\label{fig:stability-burger}
\end{figure}

If one considers not too large displacements the 
evolution of the condensate wave function can be safely described 
by a state of well defined quasimomentum (before instabilities occur) 
\cite{nesi}.
Using the formalism of Secs. III and IV we first calculate the
stability diagram predicted by the linearized GP theory for an
infinite cylinder having the same linear density as the one
of the actual 3D elongated condensate near its center. The diagram
is plotted in the upper part of Fig.~\ref{fig:stability-burger}.
In the lower part we show the center-of-mass velocity, defined
as $v_{\rm cm}=\partial E_p / \partial p$, obtained by solving 
the stationary GP equation (solid line). The two horizontal dotted 
lines indicate the critical velocities for the onset of energetic 
($v_E$) and dynamical ($v_D$) instability. 

\begin{figure}
    \centerline{\includegraphics[width=8.5cm,clip=]{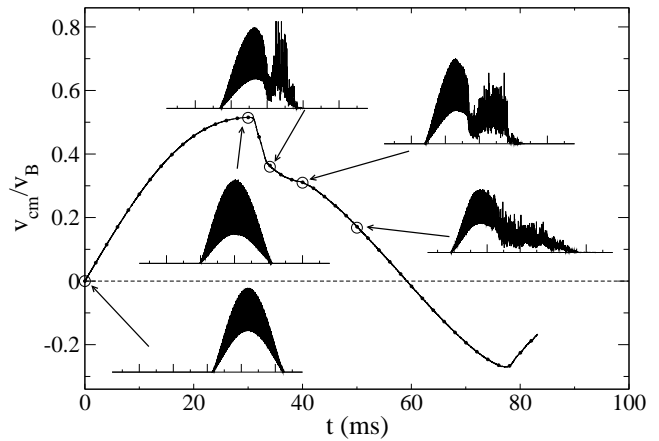}}
 \caption{Center-of-mass velocity of the condensate during the
    its evolution in the harmonic trap after an initial displacement 
    of $60$~$\mu$m. The velocity initially increases, as expected 
    for the harmonic dipole oscillation of the whole condensate, up 
    to a critical value at which the condensate breaks up. The 
    corresponding quasimomentum $p$ lies in the dynamical unstable 
    region of Fig.~\protect\ref{fig:stability-burger}. In the insets 
    we plot the linear density of the condensate along $z$ at 
    different times. The horizontal ruler has a width of $280~\mu$m
    and the center of the trap is at mid point.}
    \label{fig:vmean}
\end{figure}

Then we numerically integrate the full 3D GP equation, using the
experimental parameters. The evolution of the condensate and of 
its center-of-mass velocity, after an initial lateral displacement of 
$60$~$\mu$m, is shown in Fig.~\ref{fig:vmean}. As the figure shows, 
the condensate first accelerates towards the new trap center (to the 
left in the insets) by keeping its shape. The overall phase of the 
order parameter is also preserved, except for the addition of the 
extra phase associated with the translational motion. At a critical 
velocity the coherence is suddenly lost and the order parameter 
develops a complex structure, starting from its center. This 
critical velocity is also shown in the lower part of 
Fig.~\ref{fig:stability-burger} (horizontal dashed line). It 
turns out to be very close to the one predicted for dynamical instability 
of the cylindrical condensate and well above the threshold of
energetic instability.  Finally, the critical velocity found in our GP 
simulations nicely agrees with the velocity at which the breakdown 
of superfluidity was observed in Ref.~\cite{burger} ($\simeq0.5 v_B$), 
therefore confirming that the dynamical instability plays a major 
role in this kind of experiments, as previously argued by Wu 
and Niu \cite{wuniu}. 

\begin{figure}
     \centerline{\includegraphics[width=7.7cm,clip=]{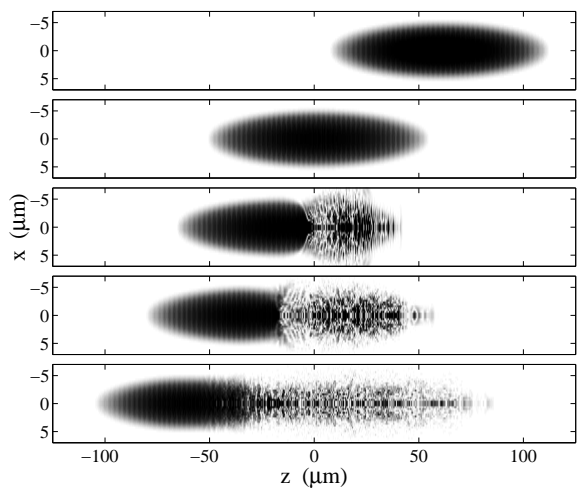}}
     \caption{Density plots in the $(z,x)$
plane for different evolution times, corresponding to the inset 
pictures of Fig.~\protect\ref{fig:vmean} ($t=0,~30,~35,~40,~50$ ms), 
after an initial displacement $\Delta z=60$ $\mu$m. The parameters are 
those of the experiment in Ref. \protect\cite{burger}.}
\label{fig:dplots}
\end{figure}

The GP simulation also reproduces the shape of the density 
distribution after the onset of the instability (see insets of
Fig.~\ref{fig:vmean} and the density plots in Fig.~\ref{fig:dplots}) 
which is characterized by a condensed/coherent part plus a  
broader incoherent distribution on the right side, as observed 
in the experiment \cite{burger}. The instability starts close
to the center of the condensate. Radial excitations are involved
in this process, as can be deduced by the occurrence of density 
patterns with radial nodes. 

By repeating the same simulations for much larger initial 
displacements we find that the system can enter well inside the 
dynamically unstable region before decoherence processes take 
place. In this case, the acceleration of the condensate is so 
fast that dynamically unstable modes have not enough time to grow 
in a significant way. Eventually, for very large displacements 
the condensate passes through the entire dynamically unstable 
region and reaches the zone boundary, at $p\sim q_B$. At this 
point, the whole condensate undergoes a Bragg reflection soon
followed by a sudden disruption process, involving solitons and 
vortex rings. This agrees with the recent GP simulations of 
Ref.~\cite{scott}, where the value $\Delta z=150~\mu$m has been 
used and the nature of this instability has been discussed in 
detail.

\section{Conclusions}

A general discussion of energetic and dynamical instabilities of 
a Bose-Einstein condensate moving in a 1D optical lattice has
been presented using the Gross-Pitaevskii theory. We have shown 
that transverse excitations are important in characterizing the
stability diagram and the occurrence of a complex radial dynamics
above the threshold for dynamical instability.

The results of the full 3D calculations have been compared with those
of an effective 1D model, the nonpolynomial Schr\"odinger equation
(NPSE), which includes the effects of the transverse direction through
a Gaussian ansatz for the radial component of the order parameter
\cite{salasnich}.  This model is shown to give accurate predictions
for the instability thresholds, that are mainly determined by the
dispersion of the lowest branch of excitations, with no radial nodes.

The linear response analysis, in combination with the direct solution 
of the time-dependent GP equation, provides a realistic framework 
to discuss the dissipative dynamics observed in recent experiments.  
As an example here we have considered the case of experiments of
Ref. \cite{burger}, providing a convincing evidence that the observed 
breakdown of superfluid flow is associated with the onset of a 
dynamical instability.

The present analysis is also relevant in connection with the more
recent experiments of Ref.~\cite{fallani} where the condensate is
loaded adiabatically in a moving lattice. In this case, interesting
results have been found by comparing the NPSE results with the
experimental data for the nontrivial dissipative behavior of 
condensates loaded in higher Bloch bands \cite{fallani}. In this
respect, our work provides a further useful information, by pointing 
out in which cases, and for which observables, the NPSE is a reliable 
approximation of the full GP theory.

\begin{acknowledgments}
We thank M. Inguscio, C. Fort, L. Fallani, L. De Sarlo, and C. Menotti
for stimulating discussions.  We are indebted to M. Kr\"amer for 
fruitful discussions on the Bogoliubov sound velocity in optical
lattices. The work has been supported by the EU under Contract 
Nos. HPRI-CT 1999-00111 and HPRN-CT-2000-00125 and by the INFM Progetto 
di Ricerca Avanzata ``Photon Matter''.
\end{acknowledgments}


\begin{thebibliography}{}

\bibitem{burger} S. Burger, F. S. Cataliotti, C. Fort, F. Minardi and
M. Inguscio, M. L. Chiofalo and M. P. Tosi, Phys.  Rev.  Lett.
\textbf{86}, 4447 (2001).

\bibitem{burger2} B. Wu and Q. Niu, Phys.  Rev.  Lett.
\textbf{89}, 088901 (2002); S. Burger \textit{et al.}, 
\textit{ibid.} \textbf{89}, 088902 (2002).

\bibitem{cataliotti} F. S. Cataliotti, S. Burger, C. Fort, P.
Maddaloni, F. Minardi, A. Trombettoni, A. Smerzi, and M. Inguscio,
Science \textbf{293}, 843 (2001).

\bibitem{lens}
F. S. Cataliotti, L. Fallani, F. Ferlaino, C. Fort, P. Maddaloni, and 
M. Inguscio, New J. Phys.  \textbf{5}, 71 (2003).

\bibitem{morsch} M. Cristiani, O. Morsch, N. Malossi, 
M. Jona-Lasinio, M. Anderlini, E. Courtade, and E. Arimondo,
Opt. Express \textbf{12}, 4 (2004).

\bibitem{oberthaler} T. Anker, M. Albiez, B. Eiermann, M. Taglieber,
 and M. K. Oberthaler, Opt. Express  \textbf{12}, 11 (2004).

\bibitem{wuniu}
B. Wu and Q. Niu, Phys.  Rev.  A \textbf{64}, 061603 (2001);
New J. Phys.  \textbf{5}, 104 (2003).

\bibitem{smerzi}
A. Smerzi, A. Trombettoni, P. G. Kevrekidis, and A. R. Bishop, Phys.
Rev.  Lett \textbf{89}, 170402 (2002).

\bibitem{menotti}
C. Menotti, A. Smerzi, and A. Trombettoni, New J. Phys. \textbf{5}, 
112 (2003).

\bibitem{pethick}
M. Machholm, C.~J.~Pethick, and H.~Smith, Phys.  Rev.  A \textbf{67},
053613 (2003); M. Machholm, A.~Nicolin, C.~J.~Pethick, and H.~Smith,
\textit{ibid.} \textbf{69}, 043604 (2004).

\bibitem{adhikari}
S. K. Adhikari, Eur. Phys.  J. D \textbf{25}, 161 (2003);
J. Phys. B: At. Mol. Opt. Phys. \textbf{36}, 3951 (2003).

\bibitem{nesi}
F. Nesi and M. Modugno, 
J. Phys. B: At. Mol. Opt. Phys {\bf 37}, S101 (2004).

\bibitem{taylor} E. Taylor and E. Zaremba, Phys. Rev. A \textbf{68}, 
053611 (2003).

\bibitem{review} F. Dalfovo, S. Giorgini, L. P. Pitaevskii, and S.
Stringari, Rev.  Mod.  Phys.  \textbf{71}, 463 (1999).

\bibitem{salasnich}
L. Salasnich, Laser Phys. \textbf{12}, 198 (2002); 
L. Salasnich, A. Parola, and L. Reatto, 
Phys.  Rev.  A \textbf{65}, 043614 (2002).

\bibitem{landau}
E.M Lifshitz and L.P. Pitaevskii, \textit{Statistical Physics} 
(Pergamon Press, Oxford, 1980), Part 2, p. 88.

\bibitem{castin}
Y. Castin, in \textit{Coherent Atomic Matter Waves}, 
edited by R. Kaiser, C. Westbrook, and F. David,
Lecture Notes of Les Houches Summer School
(EDP Sciences and Springer-Verlag, Heidelberg, 2001), pp. 1-136.

\bibitem{tozzo}
C. Tozzo and F. Dalfovo, New J. Phys. \textbf{5}, 54 (2003).

\bibitem{iyanaga} S. Iyanaga and Y. Kawada, 
\textit{Encyclopedic Dictionary of Mathematics} 
(MIT Press Cambridge, MA, 1977).

\bibitem{meret} M. Kr\"amer, C. Menotti, L. Pitaevskii, and S. Stringari,
Eur. Phys. J. D {\bf 27}, 247 (2003). 

\bibitem{zaremba} E. Zaremba, Phys. Rev. A {\bf 57}, 518 (1998); 
S. Stringari, \textit{ibid.} {\bf 58}, 2385 (1998).

\bibitem{massignan}
P. Massignan and M. Modugno, Phys.  Rev.  A \textbf{67}, 023614 (2003).

\bibitem{scott}
R. G. Scott {\em et al.}, Phys.  Rev.  A \textbf{69},  033605 (2004).

\bibitem{fallani}
L.~Fallani {\em et al.}, Phys. Rev. Lett. \textbf{93}, 140406 (2004).

\end{thebibliography}
\end{document}